\documentclass[12pt]{article}
\usepackage{amsmath, amsthm, amsfonts, amssymb,bbm}

\usepackage[english]{babel}

\title{\bf Noise induced interference fringes in trapped ultracold bosonic gases}

\author{S. Anderloni$^{a,b}$, F. Benatti$^{a,b}$, 
R. Floreanini$^{b}$, G. G. Guerreschi$^{a}$\\
\small ${}^a$Dipartimento di Fisica Teorica, Universit\`a di Trieste, 
34014 Trieste, Italy\\
\small ${}^b$Istituto Nazionale di Fisica Nucleare, Sezione di Trieste,
34014 Trieste, Italy}

\newcommand{\adag}{a^\dag}
\newcommand{\hf}{\frac{1}{2}}

\newcommand{\ket}[1]{\bigl|#1\bigr\rangle}
\newcommand{\bra}[1]{\bigl\langle#1\bigr|}
\newcommand{\op}{\mathcal O}
\newcommand{\ud}{\text d}

\newcommand{\proj}[2]{\ket{#1}\bra{#2}}
\newcommand{\id}{\mathbbm 1}
\newcommand{\hfn}{\frac{N}{2}}

\date{\null}

\begin{document}

\maketitle
 
\abstract{We study the dynamics of ultracold atoms trapped in optical double-well potentials
in presence of noise generated by an external environment.
When prepared in a Fock number state, the system shows phase coherence in the averaged density profile 
obtained using standard absorption image techniques, that disappears in absence of noise.
This effect indicates that also in many-body systems an external environment may enhance quantum coherence,
instead of destroying it.}

\section{Introduction}

Cold atoms trapped in a periodic optical potential have become a preferred test ground for the study
of quantum effects in many-body physics, {\it e.g.} in quantum phase transition,
and matter interference 
phenomena, and also for applications in quantum information (for instance, see
\cite{Lewenstein,Bloch0} and references therein).

In particular, the study of phase coherence between fractions of a same condensate
confined in a double well potential has been the focus of much theoretical and experimental analysis
as a powerful technique to obtain information about the state of a trapped condensate.
In general, phase coherence
cannot be directly observed within the trap; nevertheless, it can be put into evidence by
releasing the confining potential and letting the condensate freely expand, 
so that the atoms from the two wells may
interfere with each other: 
as a consequence, fringe patterns are formed in the
spatial density. The latter can be
obtained by illuminating the expanding clouds with a probe light
and then by collecting the corresponding absorption image.
In particular, information about the initial relative phase can be extracted from the interference
pattern \cite{Stringari}-\cite{Demler}.

One expects to observe interference fringes in the density profile only if there is phase
coherence between the wells. Instead, they appear even starting from a
state with random relative phase, like a Fock number state.%
\footnote{This phenomenon has been the focus of various investigations,
aiming to explain single shot interference formation via the detection process
(see \cite{Javanainen}-\cite{Demler} and references therein).
An alternate explanation is however possible: as further discussed in Section 3,
it makes use of a projection onto fixed-phase, coherent-like states.}
In this case however, 
the interference patterns have different offset positions in successive shots;
as a consequence, by averaging over multiple
realizations of the experiment the interference pattern disappears.

In the following, we shall study how this picture changes 
in the presence of an external environment, weakly affecting
the dynamics of the condensate in the double-well trap.
Usually, the environment is a source of noise and dissipation that are expected
to lead to decoherence.  
Instead, we shall show that, the environment may create
phase coherence between two initially uncorrelated fractions of the condensate,
an effect that may be traced to the presence
of an environment induced current between the two wells \cite{ABF}. 

We will consider a system consisting of $N$ atoms confined in a double-well trap with a very high
barrier; in absence of noise, the Fock number states with $k$ atoms in one well and $N-k$ in the other
are equilibrium states.
The presence of an environment changes the situation and makes the system an
open quantum system; its dynamics is no longer unitary and is characterized by 
the presence of noise and dissipation, so that the Fock number states are no longer stationary.
We will assume that the interaction between the atoms and the environment be weak,
with very fast decaying environment correlations.
In this case, the reduced dynamics of the trapped gas admits a Markovian ({\it i.e.} memory-less)
time-evolution described by a family of
maps forming a \textit{quantum dynamical semigroup} \cite{Gorini,Alicki,BF}.
An explicit example of such an environment has been proposed in \cite{Zoller}, where cold bosons
in an optical lattice are coupled to the Bogoliubov excitations of another condensate,
which mediates transitions between the lowest and the first excited lattice Bloch bands.
Alternatively, one may consider the coupling of the atoms to
an external classical stochastic field, with white noise correlations.

Using open quantum system techniques, we shall analyze below
the averaged density resulting from the free expansion of the gas after trap release.
We shall see that, even starting from a Fock number state, with no definite relative phase
between the states in the two wells, interference fringes may appear in the averaged density 
profile; they disappear in absence of the environment, thus showing the purely noisy
origin of the effect. We shall also see that this phenomenon may be related to the presence 
of a dissipative current between the two wells \cite{ABF}.

\section{Cold atoms in a double-well potential}

The usual description of cold atoms trapped in a double well potential
uses a two-mode Bose-Hubbard model \cite{Milburn,Jaksch} with
Hamiltonian
\begin{equation}
\label{BH}
H_{BH}=\varepsilon_1\, \adag_1a_1\,+\,\varepsilon_2\, \adag_2a_2\,+\,
U\,\Bigl[(\adag_1a_1)^2+(\adag_2a_2)^2\Bigr]\,-\,T\,(\adag_1a_2+a_1\adag_2)\ ,
\end{equation}
where $a_{1,2}$, $a_{1,2}^\dag$ annihilate and create atom states in the first, 
second well, respectively, and satisfy the Bose commutation relations $[a_i\,,\,a^\dag_j]=\delta_{ij}$.
Of the contributions to $H_{BH}$, the last one corresponds to a
hopping term depending on the tunneling amplitude $T$, the first two
are due to the trapping potential and are proportional to the depth of
the wells: in the following we will consider the case of a
symmetric trap, \textit{i.e.} $\varepsilon_1=\varepsilon_2$. Finally,
the third term, quadratic in the number operators $a^\dag_i
a_i$, takes into account repulsive Coulomb interactions inside each well.

Let $\{\ket{\phi_i}\}_{i=1}^\infty$ be a complete set of orthonormal single-particle atom states;
the creation operator $\hat{\psi}^\dag(x)$ of an atom at position $x$ 
can then be decomposed as
\begin{equation}
\label{wannier}
\hat{\psi}^\dag(x)=\sum_{i=1}^\infty\phi_i^*(x)\,a^\dag_i\ ,
\end{equation}
where $a^\dag_i$ creates an atom in the state $\ket{\phi_i}=a^\dag_i\ket{vac}$ with wavefunction
$\phi_{i}(x)=\bra{x}\phi_i\rangle$. The Bose-Hubbard Hamiltonian~(\ref{BH}) results from a
\textit{tight binding approximation}, where only the first two of the
basis vector are relevant; in this case
$\phi_{1,2}(x)$ are orthogonal functions, $\phi_1$ localized
within the first well, $\phi_2$ within the second one.

The total number $N$ of atoms is conserved by (\ref{BH}).
Therefore, the Hilbert space of the system is $N+1$-dimensional and can be
spanned by Fock number states of the form
\begin{equation}
\label{fock}
\ket{k,N-k}:=\frac{(a_1^\dag)^k(a_2^\dag)^{N-k}}{\sqrt{k!}\sqrt{(N-k)!}}\ket{vac}\ ,
\end{equation}
describing the situation in which the first well is filled with $k$ atoms,
while the other one contains $N-k$ particles. They are obtained by the action of the
creation operators on the vacuum state. These states turn out to be eigenstates
of the Bose-Hubbard Hamiltonian when the tunneling term can be neglected.

Alternatively, one can consider
the so-called phase-states
\begin{equation}
\label{SF}
\ket{\varphi,\xi;N}=\frac{1}{\sqrt{N!}}\left(\sqrt\xi e^{i\frac{\varphi}{2}}\adag_1\,+\,
\sqrt{1-\xi}e^{-i\frac{\varphi}{2}}\adag_2\right)^N\ket{vac}\ ,
\end{equation}
depending on two real parameters, an amplitude $\xi\in[0,\ 1]$ and
a phase $\varphi\in[0,\ 2\pi]$.
These states describe a physical situation where all $N$ atoms are in a same
superposition of a two-mode single particle state with relative phase $\varphi\,$;
from a many-body point of view, such states correspond to
a macroscopic wavefunction obeying the Gross-Pitaevski equation
\cite{Stringari}. Moreover, the parameter $\xi$ is related
to the mean value of the relative occupation number,
$\bra{\varphi,\xi;N}(\adag_1a_1-\adag_2a_2)\ket{\varphi,\xi;N}=N(2\xi-1)$.

These states form an overcomplete set
\begin{equation}
\label{completeness}
\mathbbm 1_N=\frac{N+1}{2\pi}\int_0^1\ud\xi\int_0^{2\pi}\ud\varphi\,
\proj{\varphi,\xi;N}{\varphi,\xi;N}\, .
\end{equation}
Although not orthogonal, for sufficiently large $N$
(a very common situation in actual experiments), the phase states become nearly so,
\begin{equation}
\label{orthogonality}
\bra{\varphi,\xi;N}\varphi',\xi';N\rangle\approx\frac{1}{N}\, \delta(\xi-\xi')\,
\delta(\varphi-\varphi')\, .
\end{equation}
One can then decompose any $N$ atom state in terms of these states;
in particular, for Fock number states, one can write
\begin{equation}
\label{kexpansion0}
\ket{k,N-k}=\frac{N+1}{2\pi}\int_0^1\ud\xi\int_0^{2\pi}\ud\varphi\,
\bra{\varphi,\xi;N}k,N-k\rangle\, \ket{\varphi,\xi;N}\ ,
\end{equation}
where the overlapping functions are explicitly given by
\begin{equation}
\label{overlap} 
\langle \varphi,\xi;N | k,N-k\rangle=
\binom{N}{k}^{1/2}\, \xi^{\frac{k}{2}}\, (1-\xi)^{\frac{N-k}{2}}\,
e^{-i\varphi\left(k-\frac{N}{2}\right)}\ .
\end{equation}
Further, the action of $\hat\psi(x)$ on the phase states can be easily derived
using (\ref{wannier}),
\begin{equation}
\hat\psi(x)\ket{\varphi,\xi;N}=\sqrt N \bigl(\sqrt\xi\, \phi_1(x)e^{i\frac{\varphi}{2}}+%
\sqrt{1-\xi}\, \phi_2(x)e^{-i\frac{\varphi}{2}}\bigr)\ket{\varphi,\xi;N-1}\, .
\end{equation}

\section{Density profile after free expansion}

As explained in the Introduction, our aim is to study the 
spatial density profile obtained by releasing the trap
and by subsequently letting the cold atoms freely
expand for a time $\tau$.
If $\tau$ is large enough, the two fractions of cold atoms
interfere with each other thus providing a spatial fringe-pattern
that one can probe by light-absorption.

Suppose the trap is released at $\tau=0$ when the cold atoms are in
the phase state (\ref{SF}). In order to describe the free expansion, disregarding atom-atom
interactions, we make use of the following model: we let the single-particle states evolve
under the action of an unitary
operator $U_\tau$; notice that, by the unitarity of $U_\tau$, the basis states 
$\ket{\phi_i}$
evolve in another set of basis states,
\begin{equation}
\ket{\phi_i}_\tau=U_\tau\ket{\phi_i}:=a^\dag_i(\tau)\ket{vac}\, ,
\qquad a^\dag_i(\tau)\equiv U_\tau a^\dag_iU_\tau^\dag\, .
\end{equation}

Thanks to the absence of interactions during the expansion, the evolution of the
phase states follows immediately from the evolution of the single particle states,
\begin{equation}
\label{SFt}
\ket{\varphi,\xi;N}_\tau=\frac{1}{\sqrt{N!}}\left(\sqrt\xi\, e^{i\frac{\varphi}{2}}
\adag_1(\tau)\,+\,
\sqrt{1-\xi}\, e^{-i\frac{\varphi}{2}}\adag_2(\tau)\right)^N\ket{vac}\ ,
\end{equation}
while both the relations (\ref{completeness}) and (\ref{orthogonality}) still
hold for any $\tau$ \cite{ABFTletter}.

The annihilation operator (which does not evolve in time) could be again
decomposed over the new basis
\begin{equation}
\label{wanniert}
\hat\psi(x)=\sum_{i=0}^\infty \phi_i(x,\tau)\, a_i(\tau)\, , \qquad \phi_i(x,\tau)
=\bra x \phi_i\rangle_\tau\ ,
\end{equation}
and therefore the action on the freely evolved phase state reads
\begin{equation}
\hat\psi(x)\, \ket{\varphi,\xi;N}_\tau=\sqrt N \bigl(\sqrt\xi\, \phi_1(x,\tau)\, 
e^{i\frac{\varphi}{2}}+
\sqrt{1-\xi}\, \phi_2(x,\tau)\, e^{-i\frac{\varphi}{2}}\bigr)\ket{\phi,\xi;N-1}_\tau\, .
\end{equation}

If the two fractions of the condensate expand independently, we can explicitly evaluate the
relative phase gained during the ballistic expansion.
Writing $\phi_i(x,\tau)=|\phi_i(x,\tau)|e^{i\theta_i(x,\tau)}$, the
phase $\theta_i(x,\tau)$ can be obtained, for large $\tau$, in terms of the
classical behavior of the velocity
$\displaystyle
\vec v=\frac{\hbar}{m}\vec\nabla\theta_i=\frac{\vec r}{\tau}$ \cite{Stringari}.
We shall focus upon the direction $x$, hence upon the phase modulation
$$
\theta_{1,2}(x,\tau)=\hf\frac{m(x-x_{1,2})^2}{\hbar\tau}\, ,
$$
where $x_i$ is
the position of the $i$-th well; it follows that the
overall relative phase after a time of flight $\tau$ can be well approximated by
\begin{equation}
\label{rel-phase-free}
\theta_1(x,\tau)-\theta_2(x,\tau)=\frac{md}{\hbar\tau}x\ ,
\end{equation}
where the trap is considered to be located along the $x$ axis,
centered at $x=0$ with the two wells at position $x_1=-d/2$, 
$x_2=d/2$, respectively.

Then, starting from a phase state, the one-particle spatial density 
explicitly reads
\begin{eqnarray}
n_{\varphi,\xi}(x,\tau)&=&{}_\tau\bra{\varphi,\xi;N}\hat\psi^\dag(x)\,
\hat\psi(x)\ket{\varphi,\xi;N}_\tau
\nonumber \\
\nonumber
&=&N\,\Biggl(\xi\, |\phi_1(x,\tau)|^2\,+\,(1-\xi)\, |\phi_2(x,\tau)|^2\,+\\
&&\hskip .5cm
\label{SFdensity}
+\,2\sqrt{\xi(1-\xi)}\, |\phi_1(x,\tau)|\,|\phi_2(x,\tau)|\,
\cos\left(\frac{md}{\hbar \tau}x\,+\,\varphi\right)\Biggr)\ .
\end{eqnarray}
The last term in this equation exhibits the presence of fringes that are
equally separated by a distance
$\displaystyle\ell\propto\frac{\hbar \tau}{md}$
with an offset position fixed by the initial phase $\varphi$.
\medskip

The situation is completely different if one starts instead from a Fock number
state; in this case the mean value of the density operator after the free expansion
can be calculated by decomposing the state using
the analog%
\footnote{Note that the scalar products 
$\langle\varphi,\xi;N|k, N-k\rangle$ in~(\ref{kexpansion}) do not depend on time.} 
of~(\ref{kexpansion0}) at \mbox{time $\tau$},
\begin{equation}
\label{kexpansion}
\ket{k,N-k}_\tau=\frac{N+1}{2\pi}\int_0^1\ud\xi\int_0^{2\pi}\ud\varphi\,
\bra{\varphi,\xi;N}k,N-k\rangle\, \ket{\varphi,\xi;N}_\tau\ .
\end{equation}

Experimentally, the
light-absorption image relative to a Fock number state
shows the same fringe pattern~(\ref{SFdensity}) of the
superfluid state, whereas one would expect no interference fringes at all.
However, if one
repeats the experiment with the same initial conditions,
the fringe patterns are obtained with randomly distributed offset positions.
Therefore, by averaging over them, the fringes cancel out and the averaged
density profile does not exhibit any interference pattern~\cite{Stringari}-\cite{Demler}.

In order to interpret this effect,
notice that the absorption image amounts to a measurement of the initial relative 
phase $\varphi$, for it gives a one-shot density profile of 
the form~(\ref{SFdensity}).
Furthermore, starting from a phase-state $\ket{\varphi,\xi;N}$, every single
repetition of the experiment will always give 
the same interference pattern $n_{\varphi,\xi}(x,\tau)$ \cite{Leggett}.
On the contrary, when the initial state at $\tau=0$ is a Fock number stare $\ket{k,N-k}$, 
each single shot will correspond to a random
selection of one of the phase-states (\ref{SFt}) in the expansion (\ref{kexpansion}),
with probability densities that are independent of the relative phase and proportional to
\begin{equation}
|\bra{\varphi,\xi;N}k,N-k\rangle|^2=\binom{N}{k}\xi^k\, (1-\xi)^{N-k}\ .
\label{weight}
\end{equation}
In other words, every single-shot measurement of the density profile
in the state \mbox{$\ket{k,N-k}_\tau$} resulting from the Fock number state $\ket{k,N-k}$,
randomly projects out the density profile
of one of the phase-states $\ket{\varphi,\xi;N}_\tau$, with weight (\ref{weight}).
As a consequence, the averaged density profile reads
\begin{eqnarray}
\label{densityNk}
n_{k}(x,\tau)&:=&
\frac{N+1}{2\pi}\int_0^1\ud\xi\int_0^{2\pi}\ud\varphi\, |\langle\varphi,\xi;N|k,N-k\rangle|^2\,
n_{\varphi,\xi}(x,\tau) \nonumber \\
&=&\frac{N}{N+2}\,
\Bigl[(k+1)|\phi_1(x,\tau)|^2\,+\,(N-k+1)|\phi_2(x,\tau)|^2\Bigr]\ ,
\end{eqnarray}
and shows no interference pattern (see \cite{ABFTletter} for further details).

The density profile at time $\tau$ after trap release depends on the state
of the $N$-atom system at time $\tau=\,0$.
Since we are interested in studying the
effects of a dissipative open dynamics during the interval of time $t$
spent by the system within the confining double-well potential,
we will describe the system state at time $t$ by a generic density matrix
$\rho_t$:
\begin{equation}
\label{ro-dec}
\rho_t=\sum_{k,q=0}^N\,
R^t_{kq}\proj{k,N-k}{q,N-q} \ .
\end{equation}
Notice that the information
about the dynamics during the time $t$ spent by the atoms within the
double-well trap is contained in the coefficients $R^t_{kq}$.

In order to construct the averaged density profile in such a case, one considers 
the ensemble of profiles
$n_{\varphi,\xi}(x,\tau)={}_\tau\bra{\varphi,\xi;N}\hat\psi^\dag(x)\hat\psi(x)
\ket{\varphi,\xi;N}_\tau$ that are obtained in many repetitions of the experiment,
with the same initial conditions.
Each image shows interference
fringes of the form (\ref{SFdensity}) and appears in the statistical
ensemble of all collected images with weights proportional to
\begin{equation}
\label{stat-mat}
\langle\varphi,\xi;N|\rho_t|\varphi,\xi;N\rangle=
\sum_{k,q=0}^N\,R^t_{kq}\,
\sqrt{\binom{N}{k}\binom{N}{q}}\xi^{\frac{k+q}{2}}(1-\xi)^{N-\frac{k+q}{2}}\,e^{-i\varphi(k-q)}\ .
\end{equation}

The resulting one-particle spatial density profile
is then given by the corresponding statistical average
\begin{eqnarray}
\label{density}
\hskip -.5cm
n_{\rho_t}(x,\tau)&=&
\frac{N+1}{2\pi}\int_0^1\ud\xi\int_0^{2\pi}\ud\varphi\, \langle\varphi,\xi|\rho_t|\varphi,\xi\rangle\,
n_{\varphi,\xi}(x,\tau) \nonumber \\
&=&\frac{N}{N+2}\,\biggl\{\sum_{k=0}^N R^t_{kk}\Bigl[
(k+1)|\phi_1(x,\tau)|^2\,+\,(N-k+1)|\phi_2(x,\tau)|^2\Bigr]
\nonumber \\
&+&\sum_{k=0}^{N-1}\,
R^t_{kk+1}\,
\sqrt{(k+1)(N-k)}\,
|\phi_1(x,\tau)|\,|\phi_2(x,\tau)|\,
e^{-i\frac{md}{\hbar\tau}x}+\ c.c.\biggr\}\ .
\end{eqnarray}
The presence of visible interference fringes is strictly related to the next-to-diagonal
entries of the density matrix $R^t_{kk+1}$ (and their hermitian conjugates).%
\footnote{
The evolution of the atoms within the trap affects only the coefficient
$R^t_{kq}$ and depends on the time $t$ spent inside the trap;
the time of flight $\tau$ refers instead to the time interval between the releasing
of the trap
and the snapshot of the density and it is a parameter of the experimental setup.}

Notice that the previous average can be computed as the mean value of the
density operator $\hat{\psi}^\dag(x)\hat{\psi}(x)$ with respect to a density matrix
$\widetilde{\rho}_t$ which results from the freely evolved state $U_\tau\,\rho_t\,U^\dag_\tau$ 
by the action of a generalized measurement,
a so-called \textit{Positive Operator Valued Measure} (POVM)
\cite{Peres,AF},
consisting of the family of projectors 
$P_{\varphi,\xi;N}:=\ket{\varphi,\xi;N}\bra{\varphi,\xi;N}$.
Explicitly, one has
\begin{equation}
\label{POVM}
\widetilde{\rho}_\tau=\frac{N+1}{2\pi}\int_0^1\ud\xi\int_0^{2\pi}\ud\varphi\,P_{\varphi,\xi;N}\,
U_\tau\,\rho_t\,U^\dag_\tau\,P_{\varphi,\xi;N}\ .
\end{equation}
The map from $\rho_t$ to $\widetilde{\rho}_t$ corresponds to the canonical description
of the quantum measurement process associated with the POVM $\{P_{\varphi,\xi;N}\}$.
One notices that it does not correspond to a measurement of the number operator
in the state $U_\tau\,\rho_t\,U^\dag_\tau$ which gives an average number of the form
\begin{eqnarray}
\nonumber
&&\text{Tr}\Big(\hat\psi^\dag(x)\hat\psi(x)\, U_\tau\rho_tU^\dag_\tau\Big)=
N\int\ud x_2\cdots x_N\bra{x,\, x_2,\dots,\, x_N}U_\tau\rho_tU^\dag_\tau
\ket{x,\, x_2,\dots,\, x_N} \nonumber \\
\nonumber
&&\hskip 3.7cm =\sum_{k=0}^N R^t_{kk}\Bigl[
k|\phi_1(x,\tau)|^2\,+\,(N-k)|\phi_2(x,\tau)|^2\Bigr]
\nonumber \\
&&\hskip 1cm +\biggl\{\, \sum_{k=0}^{N-1}\,
R^t_{kk+1}\,
\sqrt{(k+1)(N-k)}\,
|\phi_1(x,\tau)|\,|\phi_2(x,\tau)|\,
e^{-i\frac{md}{\hbar\tau}x}+\ c.c.\biggr\}\ ,
\label{density0}
\end{eqnarray}
which is often used to fit experimental data; this average corresponds to a POVM consisting of
projections onto eigenstates of the operator $\hat\psi^\dag(x)\hat\psi(x)$, not onto phase-states.

For $N$ large, the difference between (\ref{density0}) and (\ref{density}) results very small
for all practical purposes. Furthermore, the difference does not show up in the oscillating
terms. In the following we will make use of the the description of the absorption images
in terms of a POVM consisting of the projections onto the phase states,
since this appears to describe more adequately the absorption image formation \cite{ABFTletter}.

\section{Noisy dynamics within the trap}

In this Section we study the dynamics of the $N$ atoms within a trap immersed in an external
environment; in particular, we shall focus upon its effects  
on the ground state $\ket{N/2,N/2}$ of (\ref{BH}) with a very high potential barrier. 

The density matrix of the total system, atoms plus environment, $\rho_{SE}$
evolves unitarily under the action of a total hamiltonian which can be written in the form
\begin{equation}
\label{HSE}
H_{SE}=H_{BH}\otimes\id_E+\id\otimes H_E+ H_I\ ,
\end{equation}
where $H_{BH}$ is as in (\ref{BH}) and describes the motion of the system in absence of the environment, 
$H_E$ is the hamiltonian of the environment, while $H_I$
describes the interaction between them
and can be taken to have the general form
\begin{equation}
\label{interaction}
H_I=\sum_i V_i\otimes B_i
\end{equation}
where $V_i$ and $B_i$ are suitable system and
environment hermitian operators, respectively.

The evolution of the system density matrix $\rho=\text{Tr}_E(\rho_{SE})$ can be
obtained by tracing out the environment degrees of freedom from the standard
unitary evolution of the total system initial state $\rho_{SE}$.
In general this operation leads to a
master equation affected by memory terms and nonlinearities.
The latter can be avoided by choosing an initially
uncorrelated state of the system and the bath, $\rho_{SE}=\rho\otimes\rho_E$;
this is a common situation in experimental contexts when the environment is
supposed to be in a reference equilibrium state
$\rho_E$ .

On the other hand, memory effects can be considered a short time phenomenon; if we are not
interested in this transient regime, they can be neglected by studying the
effective evolution of the system on a slower timescale.
This amounts to a Markov approximation~\cite{Alicki,Gorini,BF}
which is only plausible when a neat separation exists between system and environment
timescales.
Necessary conditions for this approximation are for instance satisfied within the so called
\textit{singular coupling limit} \cite{Gorini2,Frigerio,Verri}, where the characteristic decay
times of the environment correlations,
\begin{equation}
\label{G}
G_{ij}(t)=\langle B_{i}(t)B_{j}\rangle\equiv{\rm Tr}\Big[B_{i}(t)\, B_{j}\, \rho_E\Big]\ ,\quad
B_{i}(t)=e^{itH_E} B_i\, e^{-itH_E}\ ,
\end{equation}
are very small when compared to the typical time scale 
of the system. 
This condition is satisfied by a thermal bath in
equilibrium at very high temperature or by an external stochastic white-noise
classical field; in this latter case, the matrix $[G_{ij}(t)]$ turn out to 
be real symmetric.
In \cite{Zoller}, a suitable bath fulfilling the above conditions has been proposed, where
the system is coupled with the Bogoliubov excitations of another condensate, 
which drive a transition of the trapped atoms to the first excited lattice Bloch band, 
with a characteristic transition rate $\kappa$.%
\footnote{It is interesting to notice that another possible mechanism inducing dissipation is provided by
inelastic collisions among atoms on the same lattice site, that can be tuned on by a nonvanishing
imaginay part of the interaction parameter $U$ \cite{Cirac3}. These effects would nevertheless
lead to atom loss, {\it i.e.} to non conservation of $N$,
requiring an extension of the formalism to be properly treated.}

In the singular-coupling limit, 
the dissipative time-evolution of the open system density matrix is generated by a
master equation of the form
\begin{equation}
\label{ME}
{\partial\rho(t)\over\partial t}=\mathbb L[\rho(t)]
\equiv-i[H_{BH}+H^{(2)},\ \rho(t)]+\mathbb D[\rho(t)]\ ,
\end{equation}
where the noise contributes with an effective correction $H^{(2)}$ to the free hamiltonian
and with a linear operator $\mathbb D[\cdot]$ which cannot be recast in hamiltonian form.
Their explicit expressions take the standard Kossakowski-Lindblad form \cite{Gorini,Alicki,BF}:
\begin{eqnarray}
	\label{D}
	&&\mathbb D[\rho]=\sum_{ij=1}^{4}c_{ij}\left[V_{j}^{\dag}\rho V_{i}
		-\frac{1}{2}\left\{V_{i}V_{j}^{\dag},\rho\right\}\right]\ , \\
	&&H^{(2)}=\sum_{ij=1}^{4}s_{ij}V_{i}V_{j}^{\dag}\ .
	\label{H2}
\end{eqnarray}
The operators $V_i$ are the operators involved in the
interaction hamiltonian (\ref{interaction}),
while the entries of the two hermitian matrices $[c_{ij}]$ and $[s_{ij}]$ 
embody all the information about the effects induced by the environment
and are obtained from the correlation functions of the environment:
\begin{eqnarray}
\int_0^\infty\ud t\, G_{ij}(t)&=&\hf\,  c_{ij}+i\, s_{ij}\ , \nonumber \\
c_{ij}&=&\int_{-\infty}^{+\infty}\ud t\, G_{ij}(t)\equiv \mathcal FG_{ij}(\omega)\big|_{\omega=0}\ , \\
s_{ij}&=&\frac{1}{2\pi}\mathcal P\int_{-\infty}^{+\infty}\ud \omega\, \frac{\mathcal FG_{ij}(\omega)}{\omega}\ ,
\end{eqnarray}
where $\mathcal F$ denotes Fourier transforming and $\mathcal P$ the principal value.

In particular $c_{ij}$ is known as Kossakowski matrix and turns out to be
positive definite, a key requirement to ensure the positivity of the
eigenvalues of the evolved density matrix $\rho_t$
at all times and therefore the consistency of the dynamics.
For instance, in the case of the environment proposed in \cite{Zoller}, 
the matrix $[c_{ij}]$ takes the explicit expression%
\footnote{The dissipative effects induced by inelastic atom collision discussed in \cite{Cirac3}
can similarly be described in terms of a master equation of the form (\ref{ME}); the magnitude
of the elements of the corresponding Kossakowski matrix is determined by ${\cal I}m(U)$.}
\begin{equation}
\label{example}
[c_{ij}]=
\begin{pmatrix}
2\kappa & -2\kappa & 0 & i\kappa \\
-2\kappa & 2\kappa & 0 & -i\kappa \\
0 & 0 & 0 & 0 \\
-i\kappa & i\kappa & 0 & \kappa/2
\end{pmatrix}\, .
\end{equation}

The operator $\mathbb L[\cdot]$ in (\ref{ME})
generates a family of maps $\gamma_t=e^{t\mathbb L}$, which forms a
\textit{quantum dynamical semigroup} \cite{Gorini,Alicki,BF}, 
fulfilling the forward in time composition law $\gamma_t\circ\gamma_s=\gamma_{t+s}$, $t,s\geq0$,
and describing the irreversible character of the dissipative evolution.

Quite in general, the presence of an environment will affect
both the tunneling amplitude and
the minima of the confining potential; the environment operators $B_i$ will then be coupled
to the following bilinear atom operators appearing in the system hamiltonian (\ref{BH})
$\adag_1a_1$, $\adag_2a_2$, $\adag_1a_2$
and $\adag_2a_1$, or more precisely 
to the four hermitian operators
\begin{equation}
\label{Kraus}
V_i=\left\{\adag_1a_1\, ,\,\adag_2a_2\, ,\, (\adag_1a_2+\adag_2a_1)\,
,\, i\, (\adag_1a_2-\adag_2a_1)\right\}\, .
\end{equation}
As a final remark, notice that the action of the environment ceases as a consequence of the
trap release.

\section{Effects of noise in the density profile}

As mentioned before, we shall consider an initial atom state of the
form
\begin{equation}
\rho_0=\proj{N/2,\, N/2}{N/2,\, N/2}\ .
\end{equation}
In absence of the environment and for a high potential barrier, this state
is the ground state of (\ref{BH}); as a consequence, the components $R^t_{kq}$
in (\ref{ro-dec}) are time independent and $R_{kq}=\,0$ if $k$ and $q$ differ from $N/2$.
Therefore in both (\ref{density}) and (\ref{density0}) no interference fringes appear.

In order to study the effects induced by the presence of noise, 
it is sufficient to consider the small time dynamics resulting from (\ref{ME}),
whose solution up to first order in time then reads
\begin{equation}
\label{small times}
\rho_t=\rho_0+it\, [H_{BH}+H^{(2)},\, \rho_0]+t\, \mathbb D[\rho_0]+\op(t^2)\ .
\end{equation}

Since $[H_{BH},\, \rho_0]=\,0$, only the dissipative contributions in (\ref{small times})
need to be studied.
The explicit expressions of $\mathbb D[\rho_0]$ and $i[H^{(2)},\rho_0]$
are collected in the Appendix. Using them, to first order in $t$, one finally gets 
\begin{eqnarray}
\nonumber
n_{\rho_t}(x,\tau)&=&
\frac{N}{2}\left(|\phi_1(x,\tau)|^2+|\phi_2(x,\tau)|^2\right)\, + \\
\nonumber
&+&\, t\, \frac{N^2}{4}\Biggl\{4\, \Im m(c_{34})\, \bigl(|\phi_2(x,\tau)|^2-
|\phi_1(x,\tau)|^2\bigr)\, + \\
\nonumber
&+& |\phi_1(x,\tau)|\, |\phi_2(x,\tau)|\Bigg[
\Big(2\,\Re e(s_{24}-s_{14}) + \Im m(c_{14}-c_{24})\Big) \cos\Big(\frac{md}{\hbar\tau} x\Big)\\
&+& 
\Big(2\,\Re e(s_{23}-s_{13}) + \Im m(c_{13}-c_{23})\Big) \sin\Big(\frac{md}{\hbar\tau} x\Big)
\Bigg]\Biggr\}\ .
\label{final}
\end{eqnarray}
The last two lines of the last equation contain oscillating terms which give rise
to interference fringes spaced by a distance $\ell\propto\frac{\hbar\tau}{md}$;
notice that these fringes disappear in absence of noise ({\it i.e.} when $c_{ij}=\,0=s_{ij}$).
While the interference pattern is the same as in (\ref{SFdensity})
and is essentially due to the ballistic expansion, the amplitude of these oscillations is proportional
to the entries of the matrices $[c_{ij}]$ and $[s_{ij}]$,
which depend on the strength of the environment correlation functions.

Since for large enough $\tau$ the atomic clouds coming from the two wells overlap,
{\it i.e.} $|\phi_1(x,\tau)|\simeq|\phi_2(x,\tau)|$, the magnitude of the dissipative
terms goes as $t\, N|c_{ij}|$ with respect to the standard, noise independent contribution.
Despite the large number $N$ of atoms, because the system environment coupling is weak
and thus the constants $c_{ij}$ and $s_{ij}$ are small \cite{Gorini2,Frigerio,Verri}, 
the first order expansion in (\ref{small times})
is meaningful even for sufficiently long system environment interaction times $t$.

For instance, if the Kossakowski matrix $[c_{ij}]$ is as in (\ref{example}), 
the magnitude of its entries is proportional to the
transition rate $\kappa$; for typical values of this parameter and of $N$,
$t$ can be large enough to allow a direct experimental observation of the interference fringes.
Finally, notice that if instead the system is coupled to a stochastic classical field,
the matrix $[c_{ij}]$ is real symmetric while all the entries of $[s_{ij}]$ vanishes \cite{Gorini2};
therefore, the oscillating terms in (\ref{final}) disappear; this means that classical correlations 
do not lead to any environment induced interference phenomena.

\section{Relations with dissipative current}

To better understand the physical origin of the environment induced interference pattern, 
let us consider the mean
value of the following current operator \cite{ABF}
\begin{equation}
\hat J=i(\adag_1a_2-\adag_2a_1)\, ,
\end{equation}
which describes the motion of the barycenter of the atom gas in the double well. Its mean value is clearly
zero in the initial state $\rho_0$, while, in the evolved state up to first order in $t$, it
becomes
\begin{eqnarray}
\langle\hat J\rangle_{\rho_t}&=&
N\, \Bigl(\hfn+1\Bigr)\, \Bigl[\Im m\bigl(c_{23}-c_{13}\bigr)
+2\, \Re e\bigl(s_{13}-s_{23}\bigr)\Bigr]
\end{eqnarray}
By comparing the last expression with (\ref{final}) one notices that, if the current operator has
nonzero mean value, then interference fringes can be observed; however the structure of the 
interference pattern is in general richer, as also 
the \textit{tunneling} operator \mbox{$\adag_1a_2+\adag_2a_1$} contributes to it. Indeed, one finds
\begin{eqnarray}
\langle\adag_1a_2+\adag_2a_1\rangle_{\rho_t}
&=&N\, \Bigl(\hfn+1\Bigr)\, \Bigl[\Im m\bigl(c_{14}-c_{24}\bigr)
+2\, \Re e\bigl(s_{24}-s_{14}\bigr)\Bigr]
\end{eqnarray}
This is precisely the situation that occurs for the Kossakowski matrix (\ref{example}); it
leads to a vanishing mean value
of the current operator, while instead the mean value of the tunneling operator is
proportional to $\kappa$.

\section{Outlook}

We have seen that, even in the ground state $\rho_0$ of the hamiltonian~(\ref{BH}), 
the presence of a noisy
environment may give rise to interference effects in the averaged density profile.
The periodic structure of these fringes comes from the ballistic
evolution of the gas and does not depend on the noise;
their oscillations amplitude are, however, directly proportional to the characteristic parameters
of the noise and vanish in absence of it.

This phenomenon can be related to a dissipative current between the wells,
so that the study of the interference fringes in the
averaged density profile gives information about this current.

The system we analyzed can in principle be engineered in the laboratory,
for instance using a second condensed gas as an environment as proposed in \cite{Zoller}:
the resulting interferences fringes can be detected by subtracting the density pattern in absence of noise. 
This effect can also be useful to further
understand the role of noise in double well potentials and in particular for the possibility
of driving the system out of the ground state $\rho_0$
without lowering the barrier.

Similar conclusions can be easily extended to the case of an optical lattice; in
this situation the capability of suitably engineered environment to create 
correlations between different sites
of the lattice could be practically very useful in order to create 
specific classes of correlated states.

%s\clearpage
\section*{Appendix}

By setting
$\ket{k}\equiv\ket{k,N-k}$ for sake of simplicity,
the explicit expressions of $\mathbb D[\rho_0]$ and $i[H^{(2)},\rho_0]$ read
\begin{eqnarray}
\mathbb D[\rho_0]&=&\hfn\, \Bigl(\hfn+1\Bigr)
\Biggl[
\Bigl(c_{33}+c_{44}+i\, (c_{34}-c_{34}^*)\Bigr)\, \proj{\hfn+1}{\hfn+1} \nonumber \\
&&\hskip .8 cm +\Bigl(c_{33}+c_{44}-i\, (c_{34}-c_{34}^*)\Bigr)\, \proj{\hfn-1}{\hfn-1}
-2\, (c_{33}+c_{44})\, \proj{\hfn}{\hfn}
\Biggr] \nonumber \\
&&+\Bigg\{\sqrt{\hfn\, \Bigl(\hfn+1\Bigr)}\,
\Biggl[
\frac{N}{4}\, \big(c_{13}^*-i\, c_{14}^*+c_{23}^*-i\, c_{24}^*\big)
-\hf\, \big(c_{13}^*-i\, c_{14}^*-c_{23}^*+i\, c_{24}^*\big) \nonumber \\
&&\hskip 3 cm -\frac{N}{4}\, (c_{13}-i\, c_{14}+c_{23}-i\, c_{24})
\Biggr]\, \proj{\hfn}{\hfn+1} \nonumber \\
&&+\sqrt{\hfn\, \Bigl(\hfn+1\Bigr)}\,
\Biggl[
\frac{N}{4}\, \big(c_{13}^*+i\, c_{14}^*+c_{23}^*+i\, c_{24}^*\big)
+\hf\, \big(c_{13}^*+i\, c_{14}^*-c_{23}^*-i\, c_{24}^*\big) \nonumber \\
&&\hskip 3 cm -\frac{N}{4}\, (c_{13}+i\, c_{14}+c_{23}+i\, c_{24})
\Biggr]\, \proj{\hfn}{\hfn-1} \nonumber \\
&&\hskip -0.7cm+\hf\, \sqrt{\Bigl(\hfn+2\Bigr)\Bigl(\hfn+1\Bigr)\Bigl(\hfn-1\Bigr)\, \hfn}\, (c_{44}-c_{33})
\Biggl[\, \proj{\hfn}{\hfn+2}+\proj{\hfn}{\hfn-2}\, \Biggr] \nonumber \\
&&+\hfn\, \Bigl(\hfn+1\Bigr)\Bigl(c_{33}-c_{44}+i\, (c_{34}+c_{34}^*)\Bigr)\,
\proj{\hfn+1}{\hfn-1}\, +\, h.\, c.\ \Bigg\}
\end{eqnarray}

\medskip

\begin{eqnarray}
i\, [H^{(2)},\rho_0]&=&i\, \sqrt{\hfn\, \Bigl(\hfn+1\Bigr)}\,
\Biggl\{\Biggl[\Bigl(\hfn+1\Bigr)\, s_{13}+\hfn\,  s_{13}^*+i\, \Bigl(\hfn+1\Bigr)\, s_{14}
+i\, \hfn\,  s_{14}^* \nonumber \\
&&\hskip .8 cm
+\, \Bigl(\hfn-1\Bigr)\, s_{23}+\hfn\, s_{23}^*+i\, \Bigl(\hfn-1\Bigr)\, s_{24}
+\hfn\, s_{24}^*\Biggr]\, \proj{\hfn+1}{\hfn} \nonumber \\
&&+\Biggl[\Bigl(\hfn-1\Bigr)\, s_{13}+\hfn\, s_{13}^*-i\, \Bigl(\hfn-1\Bigr)\, s_{14}
-i\, \hfn\, s_{14}^* \nonumber \\
&&\hskip .8 cm
+\Bigl(\hfn+1\Bigr)\, s_{23}+\hfn\, s_{23}^*-i\, \Bigl(\hfn+1\Bigr)\, s_{24}
-\hfn\, s_{24}^*\Biggr]\, \proj{\hfn-1}{\hfn} \nonumber \\
&&+2\, i\, \sqrt{\Bigl(\hfn+1\Bigr)\, \Bigl(\hfn+2\Bigr)}\, (s_{34}-s_{34}^*)\,
\proj{\hfn+2}{\hfn} \nonumber \\
&&-2\, i\, \sqrt{\Bigl(\hfn-1\Bigr)\, \hfn}\, (s_{34}-s_{34}^*)\, \proj{\hfn-2}{\hfn}
\Biggr\}+h.\, c.
\end{eqnarray}

\bigskip

\section*{Acknowledgements}
This work is supported by the MIUR project
``Quantum Noise in Mesoscopic Systems''.
S.A. acknowledges Eurotech s.p.a. for financial support.

\end{document}